\documentclass[dvips,twoside,10pt]{article}
\usepackage{amsmath,amssymb}
\usepackage{xspace}
\usepackage{tam07}
\title{Emergent rainbow spacetimes: Two pedagogical examples.}
\shorttitle{Emergent rainbow spacetimes}
\authors{Matt Visser}
\shortauthors{Matt Visser}
\affiliations{
School of Mathematics, Statistics, and Computer Science\\
Victoria University of Wellington, PO Box 600, Wellington, New Zealand
}
\email{matt.visser@mcs.vuw.ac.nz}

\abstract{
There is a possibility that spacetime itself is ultimately an emergent phenomenon, a near-universal  "low-energy long-distance approximation", similar to the way in which fluid mechanics is the near-universal low-energy long-distance approximation to quantum molecular dynamics. If so, then direct attempts to quantize spacetime are misguided --- at least as far as fundamental physics is concerned. 

In particular, this implies that we may have totally mis-identified the fundamental degrees of freedom that need to be quantized, and even the fundamental nature of the spacetime arena in which the physics takes place. Based on this and other considerations, there has recently been a surge of interest in the notion of  energy-dependent and momentum-dependent  ``rainbow'' geometries. Motivations for such a concept vary widely, from attempts at applying the renormalization group to cosmology in the large, through to attempts at interpreting the DSR models in terms of energy-dependent transformations on phase space. All of these models suffer from the fact that there is considerable disagreement and confusion as to what exactly an energy-dependent ``rainbow'' geometry might actually entail.

In the present article I will not discuss these exotic ideas in any detail, instead I will present two specific and concrete examples of situations where an energy-dependent ``rainbow'' geometry makes perfectly good mathematical and physical sense.  These simple examples will then serve as templates suggesting ways of proceeding in situations where the underlying physics  may be more complex.
The specific models  I will deal with are (1) acoustic spacetimes in the presence of nontrivial dispersion, and (2) a mathematical reinterpretation of  Newton's second law for a non-relativistic conservative force, which is well-known to be equivalent to the differential geometry  of an energy-dependent conformally flat three-manifold.  

These two models make it clear that there is nothing wrong with the concept of an energy-dependent ``rainbow''  geometry \emph{per se}. Whatever problems may arise in the implementation of any specific quantum-gravity-inspired proposal for an energy-dependent spacetime are related to deeper questions regarding the compatibility of that specific proposal with experimental reality.
}

\begin{document}
\maketitle

\section{Introduction}

If the physical spacetime described by Einstein's theory of gravity  is ``emergent'', and one should be aware that this is a very big ``if'', then the fundamental short-distance degrees of freedom can be \emph{radically} different from the long-distance ``emergent'' degrees of freedom~\cite{trieste, emergent}. A prime example of this type of behaviour is fluid mechanics where the short distance physics (quantum molecular dynamics) is radically different from the long-distance degrees of freedom  (density and velocity fields) which appear in the Euler and continuity equations. 
If a similar scenario holds for gravity, then, just as one cannot hope to get quantum molecular dynamics from quantizing fluid mechanics and the degrees of freedom appearing in the Euler's equation, one could not hope to get quantum gravity from quantizing Einstein's theory of gravity and the degrees of freedom appearing in the Einstein equation (the metric, or tetrad). In fact the metric (or tetrad) would lose their status as fundamental variables, being (like density and velocity fields) only defined in some ``mean field'' sense once one averages over appropriate microscopic degrees of freedom --- whatever they might turn out to be~\cite{trieste, emergent}.

We already have at least one very concrete and  specific example of such a behaviour in the ``acoustic spacetimes'' that emerge upon linearizing the equations of (non-relativistic, irrotational, barotropic, inviscid) fluid dynamics~\cite{rimfall, Unruh, Visser, LRR}, and there are a large number of more general situations in which ``analogue spacetimes'' can be constructed~\cite{shallow, normal, birefringent, broken, quasiparticle-mass, naturalness, LNP}.  Once one attempts to generalize the acoustic spacetimes to nontrivial dispersion relations, where the phase and group velocities can differ and have nontrivial energy and momentum dependence~\cite{broken, quasiparticle-mass, naturalness, LNP}, then one is very naturally lead to one specific class of incarnations of the notion of "rainbow geometry"~\cite{trieste}.   Furthermore under a plausible set of working hypotheses this class of ``rainbow geometries'' is remarkably similar to the class naturally arising from ``quantum gravity phenomenology''~\cite{trieste}.

Further afield energy-dependent rainbow geometries are currently of interest in cosmology, where a number of authors have tried to develop the notion of a scale-dependent metric within the context of a renormalization group flow on the space of all metrics. Very roughly speaking the idea is that if one averages the spacetime geometry over a cosmologically large length scale $L$, then the averaged metric $g_{ab}(x,L)$ should obey the Einstein equations for an effective scale-dependent Newton constant $G_N(L)$, at least to lowest order in a curvature expansion based upon the renormalization group~\cite{reuter,dou,periwal,cardone}. While ``running'' coupling constants are well-understood in particle physics, the general relativity community has traditionally been much more conservative when it comes to considering a ``running'' Newton constant or a ``running'' cosmological constant~\cite{Sola,Sola2,Guberina,Elizalde,Kosower}.

Even further afield, in some of the models loosely based on the ideas of ``doubly special relativity" (also called ``distorted special relativity'', and in either case abbreviated as DSR) there is a notion often called ``gravity's rainbow"  wherein at short distances (corresponding to individual elementary particles) there is a feature  that seems to imply that  different spacetime geometries are seen by particles of different energy-momentum~\cite{Magueijo,AC1,AC2,AC3,AC4,Magueijo2,Kimberly, Hossenfelder1, Hossenfelder2, Sindoni}. 

In many of these models there are a number of ambiguities (and depending on one's attitude, potentially significant controversies) that serve to confuse the situation. In this article I hope to clarify the situation somewhat by explicitly constructing a pair of elementary and pedagogically useful examples of energy-dependent ``rainbow'' geometries in  physical situations where all of the relevant physics is both well understood and straightforward. By doing so I hope ultimately to clarify the situation for the more complicated situations arising in cosmology, DSR, and quantum gravity phenomenology.

The two pedagogical models I will specifically deal with in this article are:
\begin{itemize}
\item Acoustic spacetimes with nontrivial dispersion relations~\cite{trieste}.
\item Newtonian mechanics reinterpreted as geodesic motion on an energy-dependent conformally flat 3-manifold. 
\end{itemize}
These two models will teach us slightly different things about what it means to be a rainbow geometry, and hopefully will eventually lead us to a useful abstract definition of rainbow geometry.

\section{Acoustic rainbow geometries}

The acoustic rainbow geometries are based on extensions of the following rigorous theorem~\cite{trieste, Unruh, Visser, LRR}:

\noindent{\bf Theorem:} Consider a non-relativistic  irrotational, inviscid, barotropic perfect fluid,  governed by the Euler equation, continuity equation, and an equation of state. The dynamics of the linearized perturbations (sound waves, phonons) is governed by a D'Alembertian equation 
\begin{equation}
\Delta_g \Phi =  {1\over\sqrt{-g}} \; \partial_a \left( \sqrt{-g} \; g^{ab} \; \partial_b \Phi\right) = 0
\end{equation}
where $g^{ab}$ is an (inverse) ``acoustic metric'' that depends algebraically on the background flow one is linearizing around:
\def\x{{\vec x}}
\def\d{{\mathrm{d}}}
\begin{equation}
g^{ab}(t,\x) \equiv 
{1\over \rho_0 \; c_0}
\begin{bmatrix}
   -1&\vdots&v_0^j\\
   \cdots\cdots&\cdot&\cdots\cdots\cdots\cdots\\
 v_0^i&\vdots&(c_0^2 \; \delta^{ij} - v_0^i \; v_0^j )\\
\end{bmatrix}.        
\end{equation}
Here $c_0$ is the (hydrodynamic) speed of sound given by $c_0^2 = \partial p/\partial\rho$, while $\rho_0$ is the background density, and $v_0$ is the background velocity of the fluid. \hfill $\Box$

It is important to realise that this is a rigorous theorem of abstract mathematical physics that leads to an \emph{a priori} unexpected occurrence of Lorentzian-signature spacetime in a fluid mechanical setting~\cite{trieste, Unruh, Visser, LRR}. The (covariant) acoustic metric is
\begin{equation}
g_{ab} (t,\x)\equiv 
{\rho_0 \over  c_0}
\begin{bmatrix}
   -(c_0^2-v_0^2)&\vdots&-v_0^j\\
   \cdots\cdots\cdots\cdots&\cdot&\cdots\cdots\\
   -v_0^i&\vdots&\delta_{ij}\\
\end{bmatrix},
\end{equation}
and the ``line element'' can be written as
\begin{equation}
\d s^2 \equiv g_{ab} \; \d x^a\; \d x^b =
{\rho_0\over c_0} 
\left[
- c_0^2 \; \d t^2 + (\d x^i - v_0^i \; \d t) \; \delta_{ij} \; (\d x^j - v_0^j \; \d t )
\right].
\end{equation}
The relevance to rainbow geometries comes once one replaces the hydrodynamic speed of sound $c_0$ by any generalized wavenumber-dependent notion of propagation speed. Specifically, replace the hydrodynamic speed of sound by any one of:
\begin{equation}
c_0 \to c(k^2) \to \left\{ \begin{array}{l}
c_\mathrm{phase}(k^2);\\
c_\mathrm{group}(k^2);\\
c_\mathrm{geometric}(k^2) = \sqrt{ c_\mathrm{phase}(k^2) \; c_\mathrm{group}(k^2)};\\
c_\mathrm{signal}.\\
\end{array}
\right.
\end{equation}
Here we define the signal speed by~\cite{Brillouin}
\begin{equation}
c_\mathrm{signal,1} = \lim_{k\to\infty} c_\mathrm{phase}(k^2),
\end{equation}
if we wish to focus on the propagation of discontinuities, and by
\begin{equation}
c_\mathrm{signal,2} = \max_{k} \; c_\mathrm{group}(k^2),
\end{equation}
if we wish to focus on information transfer via wave packets.

The point now is that these generalized wavenumber-dependent versions of acoustic geometry are all well-defined but \emph{distinct} and convey different information about the physics as one moves beyond the hydrodynamic region~\cite{trieste}:
\begin{itemize}
\item 
The rainbow metric based on phase velocity contains information about the dispersion relation.
\item 
The rainbow metric based on group velocity contains information about the propagation of wavepackets.
\item
The rainbow metric based on the geometric mean of group and phase velocities is in some sense the ``best'' local Lorentz approximation to the dispersion relation --- see further discussion below.  
\item
The non-rainbow metric based on signal velocity contains information about the overall causal structure.
\end{itemize}
Indeed if the signal velocity is finite then despite the complicated rainbow metric the overall causal structure is similar to that of general relativity, but with signal cones replacing light cones. If on the other hand the signal speed is infinite then the overall causal structure is similar to that of Newtonian physics, with a preferred global time. A secondary point to extract from this discussion is that, in contrast to general relativity, rainbow spacetimes are typically ``multi-metric'' with several different metrics  encoding different parts of the physics.

All in all, the present discussion is sufficient to guarantee the well-defined mathematical and physical existence of at least one wide class of rainbow geometries --- it is not necessarily true that all rainbow geometries can be put into this ``acoustic'' form, indeed the arguments in~\cite{trieste} identify at least one slightly wider class of rainbow geometries inspired by quantum gravity phenomenology --- we could simply think of replacing $\delta_{ij} \leftrightarrow h_{ij}$ in the acoustic rainbow geometries, where $h_{ij}$ is some Riemannian 3-metric.

\paragraph{The ``geometric mean'' rainbow geometry:}

The rainbow geometry based on 
\begin{equation}
c_\mathrm{geometric}(k^2) = \sqrt{ c_\mathrm{phase}(k^2) \; c_\mathrm{group}(k^2)}
\end{equation}
is perhaps more unusual than the others.  To see why this might be a useful object to consider, write the dispersion relation in the form
\begin{equation}
(\omega - \vec v \cdot \vec k)^2 =  c_\mathrm{phase}(k^2) \; k^2 = F(k^2),
\end{equation}
and now expand $F(k^2)$ as a function of $k^2$ around some convenient reference point $k_*^2$. Then
\begin{equation}
F(k^2) = F(k_*^2) + F'(k_*^2) \; [k^2 - k_*^2] + O([k^2-k_*]^2),
\end{equation}
which we can rewrite as
\begin{equation}
F(k^2) = \left\{ F(k_*^2) - F'(k_*^2) \;  k_*^2\right\} +   F'(k_*^2) \; k^2 + O([k^2-k_*]^2).
\end{equation}
But
\begin{equation}
 F'(k_*^2) = \left. {\partial(\omega^2)\over\partial(k^2)}\right|_{k_*^2} = 
  \left. {\omega\over k} {\partial\omega\over\partial k}\right|_{k_*^2} =
  c_\mathrm{phase}(k_*^2) \; c_\mathrm{group}(k_*^2). 
\end{equation}
Furthermore, we can define a ``mass term''
\begin{equation}
\omega_0(k_*^2) = F(k_*^2) - F'(k_*^2) \;  k_*^2 = 
\left\{ c^2_\mathrm{phase}(k_*^2) - c^2_\mathrm{geometric}(k_*^2) \right\} k_*^2,
\end{equation}
and so write
\begin{equation}
(\omega - \vec v \cdot \vec k)^2 =  \omega_0(k_*^2)  
+ c^2_\mathrm{geometric}(k_*^2) \; k^2 +  O([k^2-k_*]^2).
\end{equation}
If we simply truncate the expansion, writing
\begin{equation}
(\omega - \vec v \cdot \vec k)^2 =  \omega_0(k_*^2)  + c^2_\mathrm{geometric}(k_*^2) \; k^2,
\end{equation}
then this is the best-fit ``Lorentz invariant'' dispersion relation that is  tangent to the full dispersion relation at $k_*^2$. It is intriguing that it is this geometric mean velocity that seems to be governing the effective Hawking temperature in recent numerical calculations by Unruh~\cite{Unruh-recent}, which are a natural extension of his earlier work in~\cite{Unruh-dispersion}. (Though in those calculations $\omega_0(k_*^2)$ does not seem to occur in any natural manner.)

\section{The differential geometry of Newton's second law}

I will now change gear quite drastically, and in counterpoint present a somewhat unusual route from Newton's second law to Maupertuis' variational principle, ending up with an energy-dependent conformally flat three-geometry.  More standard presentations of various parts of this analysis can be found in the textbooks~\cite{Mechanix, Arnold}, and research articles~\cite{Maupertuis, Szydlowski}. 

Consider Newton's second law
\begin{equation}
\vec F = m\; \vec a,
\end{equation}
so that for a body subject to a ``conservative'' force field
\begin{equation}
m {\d^2 \vec x\over \d t^2} = -{\partial V(x)\over \partial \vec x}.
\end{equation}
Now suppose that you have good surveying equipment but very bad clocks. So you can tell \emph{where} the particle is, and its \emph{path} through space, but you have poor information on \emph{when} it is at a particular point. Can you reformulate Newton's second law in such a way as to nevertheless be able to get good information about the \emph{path} the body follows?

Since (for simplicity) we are in Euclidean geometry we can adopt Cartesian coordinates, and so we can write the distance travelled in physical space as
\begin{equation}
\d s = \sqrt{ \d\vec x \cdot \d \vec x}.
\end{equation}
Can we now find a differential equation for the unit tangent vector $\d \vec x/\d s$ (instead of the velocity $\d\vec x/\d t$)? By using the chain rule we find
\begin{equation}
{\d^2 \vec x\over \d t^2} = 
{\d s\over \d t} {\d \over \d s} \left[  {\d s\over \d t} {\d \vec x\over \d s}  \right],
\end{equation}
which implies
\begin{equation}
{\d^2 \vec x\over \d t^2} = 
\left({\d s\over \d t}\right)^2 \left[  {\d^2\vec x\over \d s^2}  \right] + {1\over 2}
\left[ {\d \over \d s}   \left({\d s\over \d t}\right)^2 \right] {\d \vec x \over \d s}.
\end{equation}
Putting this into Newton's second law
\begin{equation}
m\left({\d s\over \d t}\right)^2 \left[  {\d^2 \vec x\over \d s^2}  \right] = 
- {\partial V(x)\over \partial \vec x} - {1\over 2} m
\left[ {\d \over \d s}   \left({\d s\over \d t}\right)^2 \right] {\d \vec x\over \d s}.
\end{equation}
Now let's simplify this a little. From the conservation of energy we have
\begin{equation}
{1\over2} m  \left({\d\vec x\over\d t}\right)^2 + V(x) = E.
\end{equation}
But this means
\begin{equation}
\left({\d\vec x\over \d t}\right)^2 = \left({\d s\over\d t}\right)^2 = {2[E-V(x)]\over m},
\end{equation}
so that Newton's second law becomes (note that $m$ drops out)
\begin{equation}
{2[E-V(x)]} \; \left[  {\d^2 \vec x\over \d s^2} \right] = 
- {\partial V(x)\over \partial \vec x} - {1\over 2}
\left[ {\d \over \d s}  2[E-V(x)] \right] {\d  \vec x\over \d s} .
\end{equation}
This can be rewritten in terms of a projection operator as
\begin{equation}
{\d^2 \vec x\over \d s^2} =  {1\over2} 
\left[ \; I - {\d \vec x \over \d s} \otimes {\d \vec x \over \d s}\;  \right] \;\;
{\partial \ln[E-V(x)]\over \partial \vec x}.
\end{equation}
This completes the job of removing ``time'' from the equation of motion. One now has an equation strictly in terms of position $x$ and physical distance along the path $s$. 

But now let's go one step further and re-write this in terms of differential  geometry --- you should not be too surprised to see a three-dimensional conformally flat geometry drop out. To see this, consider a conformally flat three-geometry with metric
\begin{equation}
g_{ab} = \Omega^2(x) \; \delta_{ab},
\end{equation}
and note that the geodesic equations are (in arbitrary \emph{non-affine} parameterization)
\begin{equation}
{\d^2 x^a \over\d \lambda^2} + 
\Gamma^a{}_{bc} \;{\d x^b \over\d \lambda}\;{\d x^a \over\d \lambda}
=
f(\lambda)\; {\d x^c \over\d \lambda},
\end{equation}
where (with indices being raised and lowered using the flat metric $\delta_{ab}$) we have
\begin{equation}
\Gamma^a{}_{bc} = 
\Omega^{-1} \left\{ 
\delta^a{}_b \Omega_{,c} + \delta^a{}_c \Omega_{,b} - \delta_{bc} \Omega^{,a} 
\right\}.
\end{equation}
Now if we chose our parameter $\lambda$ to be arc-length $s$, \emph{as measured by the flat metric} $\delta_{ab}$, then
\begin{equation}
\delta_{ab} {\d x^a \over\d s}{\d x^c \over\d s} = 1,
\end{equation}
so that differentiating
\begin{equation}
\delta_{ab} {\d^2 x^a \over\d s^2}{\d x^b \over\d s} =
{1\over2} {\d\over\d s} 
\left[ \delta_{ab} {\d x^a \over\d s}{\d x^b \over\d s} \right]
= 0.
\end{equation}
But this permits us to evaluate
\begin{equation}
f(s) 
=  
\Gamma^a{}_{bc} {\d x_a \over\d s}{\d x^b \over\d s}{\d x^c \over\d s}  
= 
[\ln\Omega]_{,a}  {\d x^a \over\d s},
\end{equation}
and consequently the geodesic equation \emph{in this particular parameterization} is
\begin{equation}
{\d^2 x^a \over\d s^2} =
\left[ \; \delta^{ab} -  {\d x^a \over\d s}{\d x^b \over\d s} \; \right] \;\;
\partial_b \ln \Omega. 
\end{equation}
This is exactly the from of the equations previously derived for the path of a particle subjected to Newton's second law, provided we identify the conformal factor as
\begin{equation}
\Omega = \sqrt{E - V(x)}.
\end{equation}
That is: the paths of particles subject to Newton's second law follow geodesics of the conformally flat three-geometry defined by
\begin{equation}
g_{ab} = [E-V(x)] \; \delta_{ab}
\end{equation}
If we denote ``distance'' as measured by this conformal metric as $\ell$ then we have
\begin{equation}
\d\ell^2 = [E-V(x)]\;\d s^2.
\end{equation}
While we have completely eliminated time from the equation for the paths there is very definitely a price to be paid --- you now need to consider a separate geometry for each value of the energy. Minimizing the ``distance'' $\ell$ is what is commonly called Maupertuis' constant-energy variational principle.  (Though note that Landau and Lifshitz attribute this version of the variational principle to Jacobi~\cite{Mechanix}.) Also note that the classically  forbidden regions where $E<V(x)$ have opposite signature ($-\!\!-\!\!-$) to the allowed regions (+++) and correspond to an imaginary $\d\ell$.

\paragraph{$N$-particle systems:}

Furthermore, for $N$ coupled particles of mass $m_i$ with (\,$i\in[1..n]$\,) Newton's  second law becomes the system of equations
\begin{equation}
m_i \; {\d^2 \vec x_i \over\d t^2} = 
- {\partial V(\vec x_1,\dots,\vec x_N)\over \partial \vec x_i}.
\end{equation}
It is easy to show that the paths swept out by this system of ODEs are geodesics in a conformally flat $3N$ dimensional configuration space with metric
\begin{equation}
\d\ell^2 = 
\left\{ E - V(\vec x_1,\dots,\vec x_N)\right\} \;
\left\{  \sum_{i=1}^N m_i \; \d s_i^2 \right\},
\end{equation}
where $\d s_i$ is ordinary physical distance for the $i$'th particle. Note that we now need to explicitly keep track of individual particles masses.

\paragraph{Generalized configuration manifolds:}

In an even more general context where we have a mechanical system with $N$ degrees of freedom, that has a kinetic energy quadratic in velocities, we may write
\begin{equation}
\mathcal{L} = {1\over2} m_{ij}(q^k) \; \dot q^i \dot q^j -  V(q^k).
\end{equation}
Here $m_{ij}(q)$ is the configuration-dependent ``mass matrix'', and the $q^k$ are generalized coordinates (living in the configuration manifold $\mathcal{M}$) that do not need to have the physical interpretation of being particle positions. A similar analysis to the above now yields an energy-dependent geometry
\begin{equation}
\d\ell^2 = g_{ij}(q^k) \; \d q^i \; \d q^j = \left\{ E - V(q^k)\right\} \; m_{ij}(q^k) \; \d q^i \; \d q^j,
\end{equation}
that lives on the classically accessible submanifold, 
\begin{equation}
\mathcal{M}(E) = \left\{ q^k : E > V(q^k)\right\},
\end{equation}
of the original configuration manifold $\mathcal{M} = \mathcal{M}(\infty)$.

\paragraph{Geodesic deviation:}

Now consider two initially parallel curves, in the 1-particle case, both corresponding to individual particles of energy $E$, separated by $\Delta x^a(s)$. Then
by considering the difference of two geodesic equations we see that their separation grows according to the rule
 \begin{equation}
{\d^2 \Delta x^a \over\d s^2} = {1\over2}
\left[ \delta^{ab} -  {\d x^a \over\d s}{\d x^b \over\d s} \right] \;\;
\partial_b \partial_c \ln [E-V(x)] \;\; \Delta x^d(s).
\end{equation}
But this is now easily turned into a statement about the focussing effect of the Riemann tensor, 
\begin{equation}
{\d^2 \Delta x^a \over\d s^2} = 
R^a{}_{bcd} \; {\d x^b\over \d s} \; {\d x^d\over \d s}  \; \Delta x^c(s),
\end{equation}
with the Riemann tensor for the conformally flat 3-geometry being
\begin{equation}
R_{abcd} =   - 2 \left\{ g_{a[d} \; R_{c]b} + g_{b[c} \; R_{d]a} \right\} -  R \; g_{a[c} g_{d]b} .
\end{equation}
The Ricci tensor is given by
\begin{equation}
R_{ab} =  {1\over2} {\partial_a \partial_b V\over E-V}  
+ {3\over4} {\partial_a V \; \partial_b V\over (E-V)^2} 
+ \delta_{ab} \left\{ {1\over2} {\nabla^2 V\over E-V} 
+ {1\over 4} { \partial_c V \; \partial^c V\over (E-V)^2 }\right\},
\end{equation}
while the Ricci scalar is
\begin{equation}
R = 2 {\nabla^2 V\over E-V} 
+ {3\over 2} { \partial_c V \; \partial^c V\over (E-V)^2 }.
\end{equation}
The formalism has straightforward generalizations to multi-particle systems and mechanical systems with general quadratic kinetic energies.

\paragraph{Optical--mechanical analogy:}

There is of course a deep relationship between the differential geometric formulation above, Fermat's principle of minimum time, and the optical-mechanical analogy often used in situations where index-gradient methods are useful. If the refractive index is a function of position, then it is well known that the path of a light ray can be determined by minimizing the optical distance
\begin{equation}
\ell = \int n(x) \; \d s.
\end{equation}
This is certainly equivalent to dealing with a conformally flat 3-geometry but the major difference here is that there is no simple way to turn Fermat's principle \emph{directly} into a energy-dependent differential geometry, at least not without going all the way back to Newton's equations to cook up a specific mechanical system that effectively mimics the refractive index $n(x)$.   This is often the main task in developing a specific instance of  the optical-mechanical analogy~\cite{Maupertuis,Maupertuis2,Nandi,Nandi2,Marklund}.

\section{Discussion}

The acoustic rainbow geometries and the  differential geometric version of Newton's second law have the great virtue that all relevant physics is clearly completely under control.  Thus they provide an existence proof for the notion of an energy-dependent geometry, without the additional theoretical complications inherent in the running cosmology, DSR, or quantum gravity phenomenology frameworks. When it comes to the notion of a scale-dependent running metric in cosmology, or an energy-dependent metric in DSR, or in quantum gravity phenomenology, these present toy models make it clear that it is not the  kinematics of single-particle motion in an energy-dependent spacetime that is in any way questionable.  The technical difficulties lie at a deeper level. Specifically:
\begin{itemize}
\item Multi-particle kinematics: In the Newton law scenario multi-particle kinematics was best dealt with by going to a $3N$ dimensional configuration space. This would be extremely unnatural in the context of general relativity, both mathematically and physically. Observationally the various E\"otv\"os-inspired experiments now verify the universality of free fall to at least one part in $10^{13}$. The standard way of building this observational fact into general relativity [and any plausible extension of general relativity] is via the Einstein equivalence principle wherein one explicitly enforces ``one spacetime for all individual particles'', not one (nonlocal?) spacetime based on the configuration space. While as we have seen in the current article, there is nothing particularly radical in the proposal of using an energy-dependent geometry \emph{per se},   there are real and fundamental experimental issues that must be addressed when it comes to positing an energy-dependent extension for general relativity. 
\item Geometrodynamics: There is a whole additional level of complexity that comes in to play when one wishes to ascribe a spacetime dynamics to the energy-dependent geometry. In DSR-inspired ``gravity's rainbow'' scenarios~\cite{Magueijo} this would seem to require an extension of the Einstein equations (and indeed an extension of the entire notion of spacetime curvature) onto the entire tangent bundle, typically with the individual tangent spaces becoming curved manifolds in their own right.  In ``running cosmology'' scenarios there is the delicate question of exactly what renormalization point to pick when solving the FRW equations for the universe as a whole. Should we (self-consistently?) pick the Hubble scale $L=c/H_0$? Or the scale defined by the space curvature? Or something else?
\end{itemize}
It is these harder problems that must be confronted when trying to make sense of more complicated specific models built on the notion of an energy-dependent spacetime.

Independent of the questions raised by these more complicated models, the energy-dependent acoustic geometries and energy-dependent conformal geometry implicit in Newton's second law are of interest in their own rights, both as a simple illustration of the mathematical formalism of differential geometry, and as a different and unusual way of looking at wave propagation and  classical mechanics.


\section*{Acknowledgments}

This Research was supported by the Marsden Fund administered by the Royal Society of New Zealand.   I also particularly wish to thank SISSA/ ISAS (Trieste) for ongoing hospitality. Various versions of the ideas in this article were also presented at the following conferences:  ``From quantum to emergent gravity: theory and phenomenology'',  June 2007, SISSA/ISAS, Trieste, Italy; at  "Enrageing ideas'', September 2007, Utrecht University, the Netherlands;  and at ``Experimental search for quantum gravity'', November 2007, Perimeter Institute, Canada. 



\end{document}